\input{aipcheck}

\documentclass[
    ,final            
  ]
  {aipproc}

\layoutstyle{6x9}

\begin{document}

\title{DVCS with CLAS}

\author{Elton S. Smith\\ for the CLAS Collaboration}{
  address={Thomas Jefferson National Accelerator Facility,\\
           12000 Jefferson Avenue, Newport News, VA 23606, USA}
}

\begin{abstract}
Generalized parton distributions provide a unifying framework for the
interpretation of exclusive reactions at high $Q^2$. The most promising
reaction for the investigation of these distributions is the hard
production of photons using Deeply Virtual Compton Scattering (DVCS). This
reaction can be accessed experimentally by determining the production
asymmetry using polarized electrons on a proton target. Pioneering
experiments with CLAS and HERMES have produced the first measurements of
this asymmetry. We will review the current experimental program to study
DVCS at Jefferson Lab. Recent high statistics data taken with CLAS at 5.75
GeV allows us to determine this asymmetry at low -$t$ in the valence region
($x_{B}$=0.1-0.5) up to a $Q^2$ of 4 GeV$^2$/$c^2$.
\end{abstract}

\maketitle


\section{INTRODUCTION}

Historically electron scattering experiments have focussed either 
on the measurements of form factors using exclusive processes or on measurements
of inclusive processes to extract deep inelastic structure functions.
Elastic processes measure the momentum transfer dependence of the form factors,
while the latter ones probe the quark's longitudinal momentum
and helicity distributions in the infinite momentum frame.
Form factors and deep inelastic structure functions measure two different
one-dimensional slices of the proton structure. While it is clear that the two 
pictures must be connected, a common framework for the interpretation
of these data have only recently been developed using  
Generalized Parton Distribution (GPD) functions 
\cite{ref:Muller,ref:Ji,ref:Radyushkin}.
The GPD's are two-parton correlation functions that encode both
the transverse spatial dependence and the longitudinal momentum
dependence. At the twist-2 level, for each quark species there are two
spin-dependent GPD's, \~E($x$,$\xi$,$t$) and \~H($x$,$\xi$,$t$), and
two spin averaged GPD's, E($x$,$\xi$,$t$) and H($x$,$\xi$,$t$).
The four GPD's are each functions of the longitudinal
momentum fraction $x$, the longitudinal momentum transfer $\xi$, and
the four-momentum transfer $t$.
The first $x$ moment of the GPD's links them to the proton's form factors,
while at $t$=0, the GPD's H and \~H reduce to the quark longitudinal 
momentum $q(x)$ and the quark helicity distributions $\Delta q(x)$,
respectively.  
The new physics which can be accessed
using exclusive reactions is contained in the dependence on $\xi$ and $t$. 
The $t$ dependence is directly related to the distribution of the parton densities 
as a function of impact parameter \cite{ref:Burkardt}.
Mapping out the GPD's will allow, for the first time, to 
obtain a 3-dimensional picture of the nucleon \cite{ref:Belitsky}.

\begin{figure}[t]
\includegraphics[width=0.8\textwidth,bb=40 90 400 280]{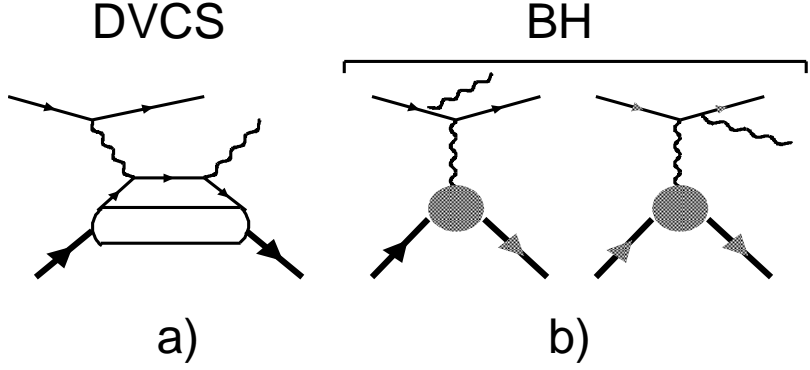}
\caption{\label{fig:handbag}
a) Deeply virtual Compton scattering (DVCS) and 
b) Bethe-Heitler processes contributing to 
$ep \rightarrow ep \gamma$ scattering.}
\end{figure}

\section{Deeply Virtual Compton Scattering}
Deeply Virtual Compton Scattering (DVCS) is one of the key reactions
to determine the GPD's experimentally, as it is the simplest process
that can be described in terms of the GPD's. 
This reaction is also expected to enter the 
Bjorken regime at relatively low photon virtuality compared to
exclusive meson production. Fig.\,\ref{fig:handbag} shows two contributions
to the production of $ep \rightarrow ep \gamma$ events.
At present CEBAF energies, the reaction is dominated by the
Bethe-Heitler (BH) amplitudes.
However, the interference term between DVCS and BH can be probed using
polarized electron beams. The interference term is given by
\begin{eqnarray}
\label{eq:int}
{\cal I} & = & {\cal T}_{DVCS}^*{\cal T}_{BH}~+~{\cal T}_{DVCS}{\cal T}_{BH}^*\, ,
\end{eqnarray}
Only the imaginary part of the DVCS amplitude survives in the
single beam asymmetry $A_{LU}$ which is accessible experimentally.
In this case, the small DVCS amplitude which depends on the GPD's, is 
amplified by the larger but well-known BH amplitude.
The interference term is dominated by the $\sin{\phi}$ moment,
where $\phi$ is  the angle between the $\gamma^*\gamma$ plane and
the electron scattering plane. At this level of approximation, the interference
term is a linear combination of Compton Form Factors ($\mathcal{H,\widetilde{H}, E}$)
\begin{eqnarray}
\label{eq:sin}
A_{LU}(\phi) & \propto & \Im m\left ( F_1 {\cal H}~+~\frac{x_{\rm B}}{2 - x_{\rm B}}(F_1 + F_2)
\widetilde {\cal H}~-~\frac{\Delta^2}{4M^2} F_2 {\cal E} \right ) \sin{\phi}
\end{eqnarray}
where $F_1(t)$ and $F_2(t)$ represents the elastic Dirac and Pauli
form factors, respectively. The Compton Form Factors are related to
the corresponding GPD's via convolution integrals \cite{ref:dvcs}.
Thus the magnitude of the azimuthal asymmetry probes a linear combination
of GPD's.

\newpage
{\vspace*{-7cm} \bf $Q^2 (GeV/c)^2$}\\*[4cm]
{\hspace*{5.5cm} \bf $x_B$ \hspace*{3.5cm} $-t~(GeV/c)^2$}\\*[0.1cm]
\begin{figure}[t]
\hspace*{1cm}
\includegraphics[width=0.7\textwidth,bb=20 150 540 660]{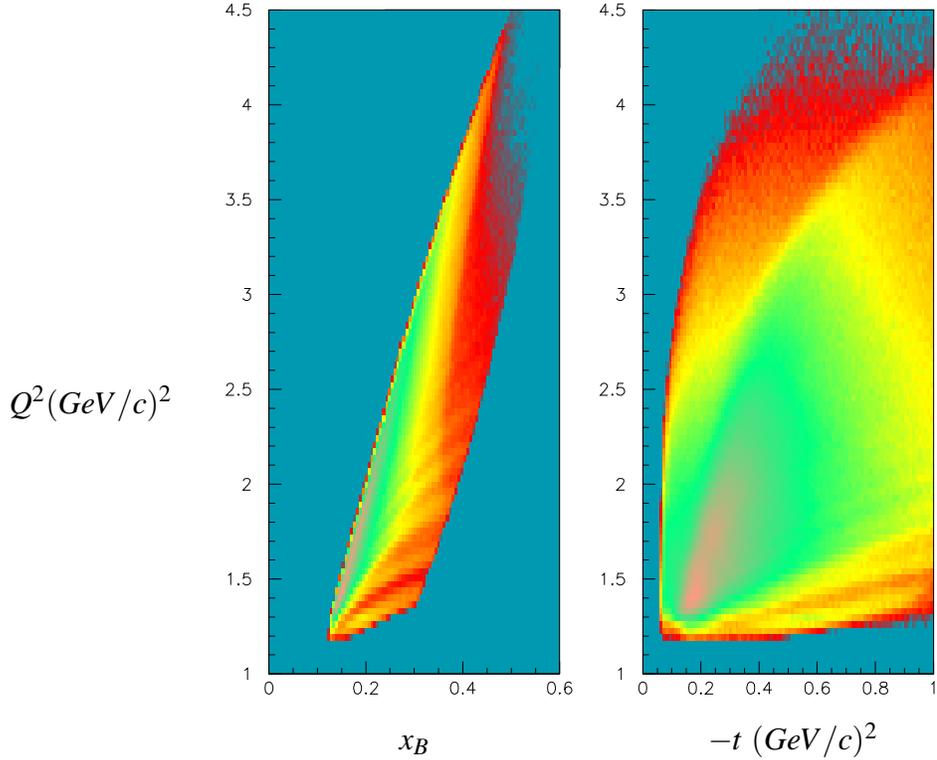}
\caption{\label{fig:e16kin}
Kinematic coverage of data taken at 5.75 GeV showing a) $Q^2$ vs. $x_B$ 
and b) $Q^2$ vs. $-t$.}
\end{figure}

\begin{figure}[t]
\begin{minipage}{7.5cm}
\includegraphics[height=8cm,clip=true]{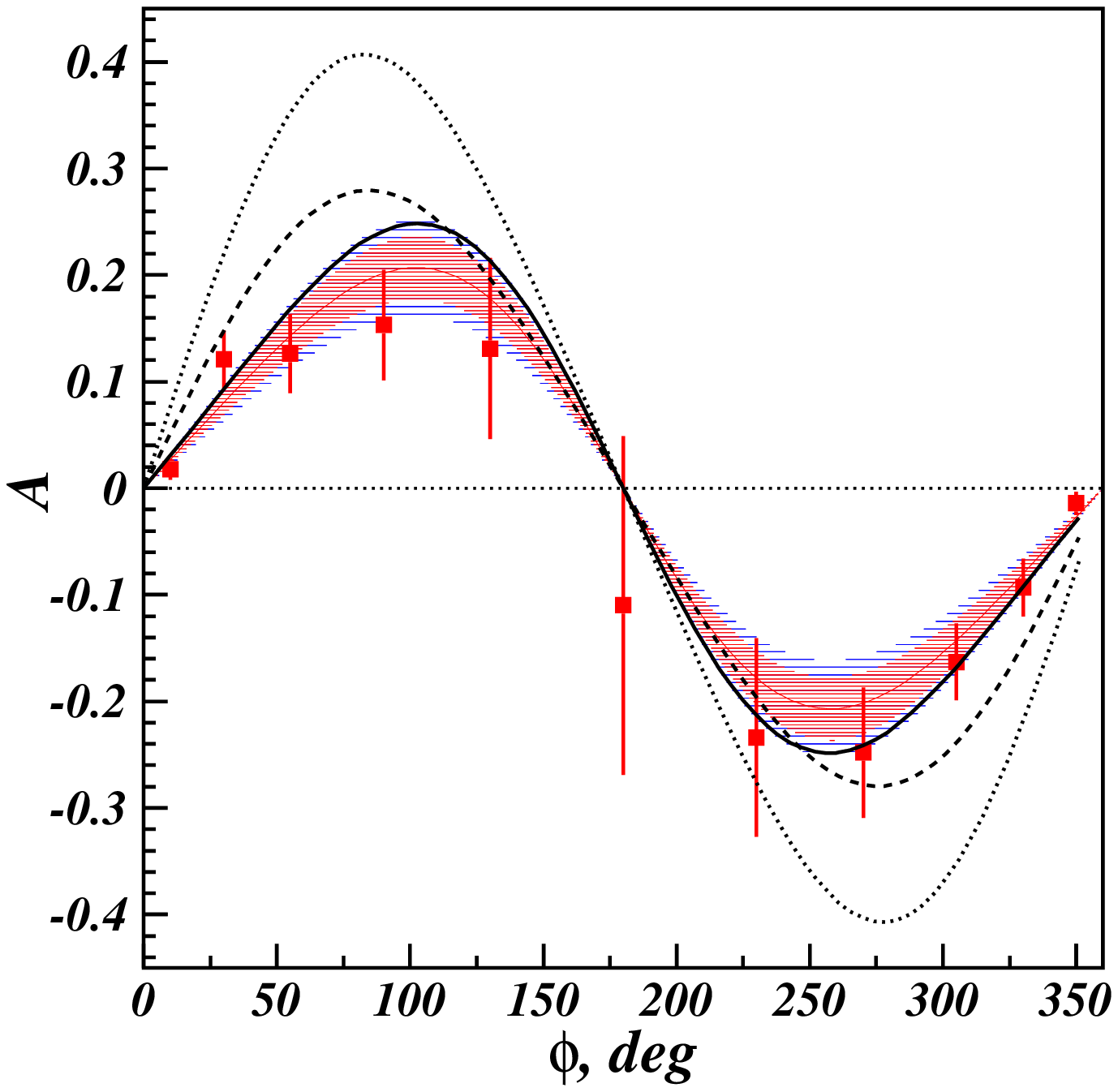}
\end{minipage}
\hfill
\begin{minipage}{7.5cm} 
\includegraphics[height=7.5cm,clip=true]{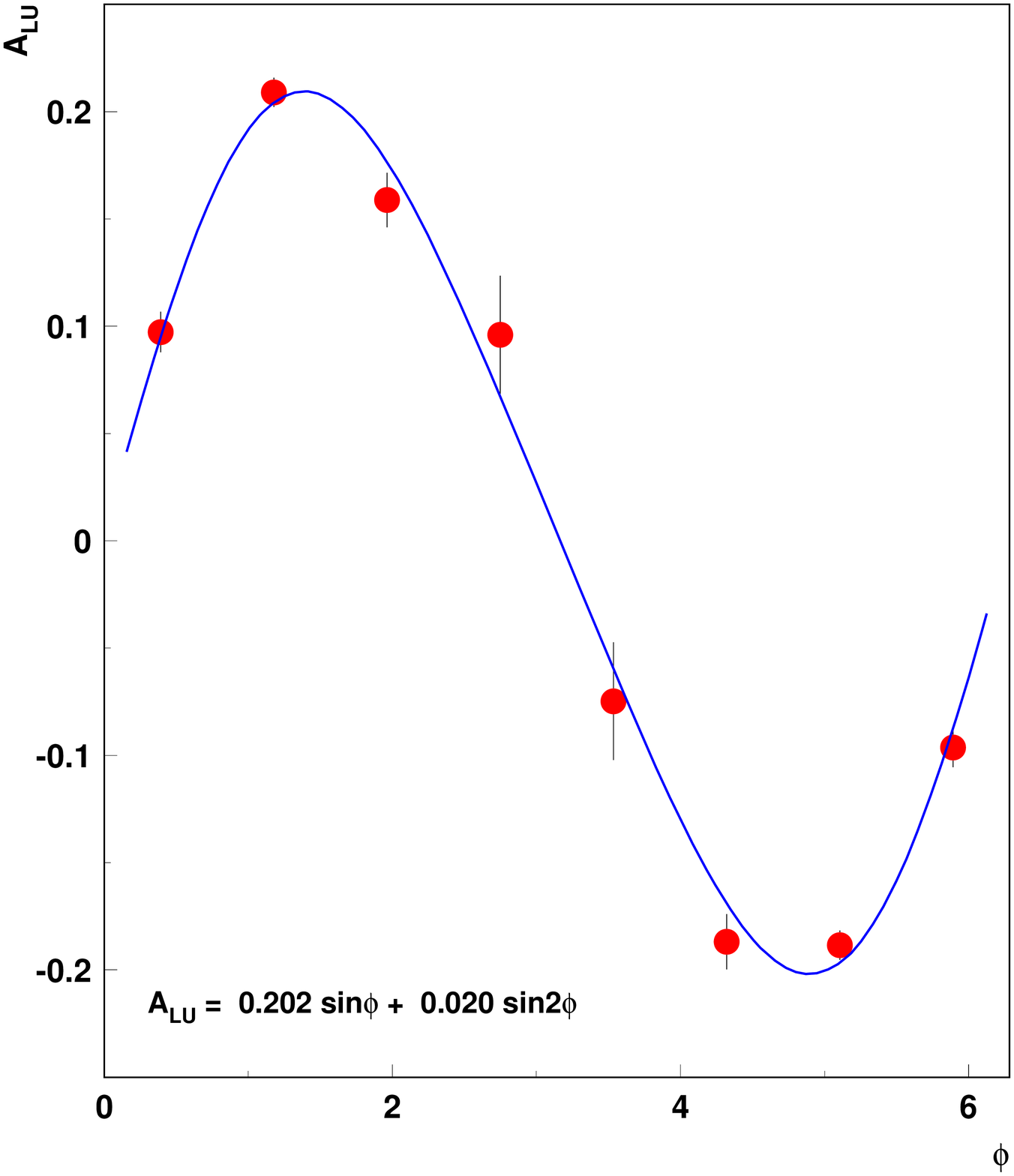}
\caption{\small{The beam spin asymmetry as a function of azimuthal angle for
a) the published data at beam energy of 
4.25 GeV and b) the new data sample at 5.75 GeV averaged over the entire kinematic range of the data. The
increased statistics available at the higher energy will allow mapping out the asymmetry
as a function of $Q^2$, $x_B$ and $t$.} 
\label{fig:asym}}
\end{minipage}
\hfill
\end{figure}

\section{Data Sample and Event Selection}
Preliminary results are now
available from an extensive data set taken with the CLAS detector \cite{ref:clas},
which was taken at a beam energy of 5.75 GeV. 
The data set corresponds to an integral
luminosity of $2.6 \times 10^{40} cm^{-2}$ accumulated on a 5-cm-long
unpolarized hydrogen target. The data were taken at a typical luminosity of
$0.9 \times 10^{34} cm^{-2} s^{-1}$ and the average 
electron beam polarization was 70\%.
The kinematic coverage for W $>$ 2 GeV is shown in Fig.\,\ref{fig:e16kin} with
$x_B$ ranging from 0.1 to 0.5 and $Q^2$ up to 4 GeV$^2$/$c^2$ at
large $x_B$. The data also covered a large range in momentum 
transfer $-t$, but analysis of DVCS was restricted to 
$-t <$ 0.5 GeV$^2$/$c^2$.

Events are selected with positive identification of the electron
in the Cerenkov counter and electromagnetic calorimeter (EC), and the
proton by time-of-flight and momentum reconstruction. The ideal experiment would also 
detect the scattered photon, but the EC has
limited coverage down to
only 8$^{\circ}$ and therefore generally misses the scattered
photon from the DVCS process. The missing mass resolution of CLAS
cannot cleanly separate the $ e p \pi^0$ from $ e p \gamma$ reactions,
and in previous analyses we have extracted the yield by fitting
the missing mass distribution to the sum of those two distributions \cite{ref:Stepanyan}.
At higher energies, this procedure becomes more difficult and 
we use the following selection criteria to obtain a clean
sample of $ e p \gamma$ events: veto photons from $\pi^0$ decay in the EC, select data
at low $-t$ using the lab angle of the scattered photon relative to the virtual photon
$\theta_{\gamma^*\gamma} \le$ 0.12 rad, and require that the missing
mass squared M$_X^2 \le$ 0.025 GeV$^2$ \cite{ref:Avagyan}.

The beam spin asymmetry at 5.75 GeV averaged over the entire data
sample is shown in Fig.\,\ref{fig:asym} compared to our published
data at 4.25 GeV \cite{ref:Stepanyan}. The beam spin asymmetry is
well described by a $\sin{\phi}$ distribution, which is expected if
the handbag diagram dominates. The small $\sin{2\phi}$ component
is of order 10\%, and can be used to place limits on the size of
twist-3 contributions. The increased statistical
power of the new event sample is reflected in the size of error bars.
This will allow us to map out the dependence of the asymmetry as a 
function of $x$ and $-t$ up to a $Q^2$ of about 4 (GeV/c)$^2$.

\section{Future Prospects}
The data presented so far is only the first step toward
probing GPD's at Jefferson Lab. There are two dedicated DVCS experiments
approved to run using 6 GeV polarized electron beams \cite{ref:Latifa}.
The first experiment E00-110 is scheduled to be run in Hall A
\cite{ref:E00-110} and is designed to determine both beam
spin asymmetries and cross section differences in three $Q^2$ intervals,
for fixed values of $x_B$. The second experiment, to take data with
CLAS, will use dedicated apparatus to over-determine the reaction
kinematics  \cite{ref:E01-113}. This experiment is building a
a forward calorimeter to directly
detect the scattered photon and a solenoidal magnet to improve the shielding of
M{\o}ller backgrounds from the target to achieve higher luminosity.
The study of GPD's to higher precision and higher $Q^2$ will continue
as one of the main research programs driving the CEBAF upgrade to 12 GeV
\cite{ref:clas++}.

\begin{theacknowledgments}
I would like to thank L. Elouadrhiri, H. Avagyan, M. Gar\c{c}on for
useful discussions and suggestions. 
This work was supported in part by the U.S. Department of Energy, including 
DOE Contract No. DE-AC05-84ER40150.
\end{theacknowledgments}


\bibliographystyle{aipproc}   


\IfFileExists{\jobname.bbl}{}
 {\typeout{}
  \typeout{******************************************}
  \typeout{** Please run "bibtex \jobname" to obtain}
  \typeout{** the bibliography and then re-run LaTeX}
  \typeout{** twice to fix the references!}
  \typeout{******************************************}
  \typeout{}
 }

\end{document}